\documentstyle[12pt]{article}

\newcommand{\be}{\begin{equation}}
\newcommand{\ee}{\end{equation}}
\newcommand{\bea}{\begin{eqnarray}}
\newcommand{\eea}{\end{eqnarray}}
\newcommand{\ep}{\varepsilon}
\newcommand{\Int}{\int\limits}

\newcommand{\nn}{\nonumber}

\newcommand{\Li}[2]{{\mbox{Li}}_{#1}\left(#2\right)}

\newcommand{\bbibitem}[1]{\bibitem{#1}}

\begin{document}

\thispagestyle{empty}
 \begin{flushright}
 INLO--PUB--21/95\\[2mm]
 hep-ph/9602396\\[9mm]
 February 1996
 \end{flushright}
\vspace{30 mm}
\begin{center}
 {\bf \large
 Small-threshold behaviour of two-loop self-energy diagrams: \\[2mm]
 two-particle thresholds}
  \footnote{This research is supported by the EU under contract
   number INTAS-93-0744 and by the Stichting FOM.}
\vspace{10 mm} \\
  F.A.~Berends$^{a,}$
  \footnote{E-mail address: berends@rulgm0.leidenuniv.nl}, 
  \ A.I.~Davydychev$^{a,b,}$
  \footnote{E-mail address: davyd@theory.npi.msu.su}
  \ and \ V.A.~Smirnov$^{a,b,}$
  \footnote{E-mail address: smirnov@theory.npi.msu.su}
\vspace{10 mm} \\
$^{a}${\em 
  Instituut-Lorentz,
  University of Leiden, \\
  P.O.B. 9506, 2300 RA Leiden, The Netherlands}
\vspace{3 mm} \\
$^{b}${\em
  Institute for Nuclear Physics,
  Moscow State University, \\
  119899 Moscow, Russia}
\end{center}
\vspace{12 mm}
\begin{abstract}
The behaviour  of two-loop two-point diagrams
at non-zero thresholds corresponding to two-particle cuts
is analyzed.
The masses involved in a cut and the external momentum
are assumed to be small as compared to some of the other masses
of the diagram.
By employing general formulae of asymptotic
expansions of Feynman diagrams in momenta and masses, we construct 
an algorithm to derive analytic approximations
to the diagrams. In such a way, we calculate several
first coefficients of the expansion.
Since no conditions on relative values of the small masses
and the external momentum are imposed, the threshold 
irregularities are described analytically. 
Numerical examples, using diagrams occurring in the
Standard Model, illustrate the convergence of the expansion
below the first large threshold.
\end{abstract}
\newpage
\setcounter{page}{2}
\setcounter{footnote}{0}

\section*{}
\begin{center}
{\bf 1. Introduction}
\end{center}

In recent years many precision calculations have been
performed in the Standard Model (SM), which are required
to match the impressive experimental accuracy. A good 
overview of the present status has been recently given
\cite{bardin}. It is to be expected that even more refined
evaluations of loop corrections in the SM and extensions
thereof will be needed in the future. Since one-loop
corrections in the SM are well established the focus will
be on two-loop studies. As physics examples one should
mention the quantity $\Delta r$  occurring in the Fermi constant
$G_F$, small angle Bhabha scattering and the $Z\rightarrow b 
\bar{b}$ decay  rate.

Such calculations are cumbersome in many respects e.g.
through the number of diagrams, the involved tensor
structures in diagrams and the complexity of 
basic building blocks i.e. scalar two-loop
diagrams\footnote{The tensor decomposition of two-loop
self-energy diagrams is described in ref.~\cite{Weigl}.}.

It is the purpose of this paper to contribute to the
techniques for calculation of the scalar two-loop diagrams.
When all masses in the propagators are non-vanishing there
exist arguments \cite{sch} that they cannot be expressed anymore
in terms of known functions like polylogarithms. 
As an example, we can mention the simplest case of a two-loop 
scalar diagram i.e. a self energy with three massive propagators, 
the so-called ``sunset'' (alias ``sunrise'')
diagram, which can be expressed in terms of Lauricella functions
\cite{berends1} or one-fold integrals \cite{GvdB,Lunev},
but no simpler results are available so far.

In such a situation two main strategies can be chosen. One
is a numerical approach, another an approximative analytical
method. Most progress has been made for self-energy diagrams.
Obviously they are the simplest ones, but they are physics-wise
also very relevant.

As to the numerical strategy one still has a number of
options. There exist a two-dimensional integral representation
\cite{Kreimer} (see also in \cite{CzKK}) originally introduced 
for the master diagram (Fig.~1a), but later on extended to all 
(also divergent) two-loop
self-energies \cite{berends2}. Another way is the use of
dispersion integrals \cite{berends1,bauberger}, whereas
other methods rely on higher dimensional integration
techniques \cite{japan,GvdB}.

The analytical approach uses explicit formulae for the
asymptotic expansion of Feynman diagrams in momenta and masses and is
based on general theorems on asymptotic expansions  \cite{as-ex} 
(see also \cite{Smirnov} for review). Some recent physical examples
of application of these formulae can be found e.g. in  \cite{appl}. 
For two-loop self-energy diagrams with general masses, expansions 
in different regions were systematically 
examined in refs.~\cite{DT1,DST,BDST}\footnote{The small momentum
expansion in the three-point case was studied in ref.~\cite{FT}
where also the techniques of conformal mapping and Pad\'e 
approximations were employed to get numerical results beyond the
threshold(s).}.

When a few terms
of the series give an adequate description this approach will
be faster than the purely numerical integration method. 
Moreover, analytic expressions are much more convenient
to deal with when the values of some masses are not fixed
(e.g. the Higgs mass) or have rather wide error bars (e.g. the
top quark mass).

The first results of the analytical approach have been obtained
outside the particle thresholds: a small momentum expansion
below the lowest threshold \cite{DT1} and a large momentum expansion
above the highest threshold \cite{DST}\footnote{For some special
cases (when some of the internal lines are massless), the expansions 
and exact results were presented in \cite{special,Tbilisi}.}. 
The region between the thresholds
poses a problem since both small and large momentum expansions
do not describe the behaviour between the  
lowest and the highest physical thresholds. 
In a previous paper \cite{BDST} it was shown that in the special
case of a zero mass threshold the small momentum expansion
could be extended to the lowest non-vanishing threshold. 
In this case the expansion coefficients involve the
zero-threshold cut which appears as powers of $\ln(-k^2)$
(where $k$ is the external momentum). 

It is the purpose of the present paper to consider 
cases when one (or two) two-particle threshold(s) is (are) small 
with respect to
the other thresholds, but not anymore zero. By using 
asymptotic expansions  in the large mass limit
it becomes possible to find a series converging above the
small two-particle threshold. The expansion coefficients
now contain the two-particle cut(s) associated with the small
threshold(s) and therefore the non-regular behaviour around the 
threshold(s)
will be exactly described. Thus the analytic approach now
is substantially extended into regions, which were hitherto
inaccessible due to the inapplicability of the used series. From
numerical examples it can be seen that the convergence holds 
beyond the small particle threshold(s). We note that there are 
no exact results (except for the zero-threshold
limit) available for the cases we are going to describe here.

The actual outline of the paper is as follows. In section~2
different small-threshold configurations are considered and
subgraphs contributing to the asymptotic expansions 
of the corresponding diagrams are described. 
In section~3 we present some analytical results 
for the lowest terms of the expansions.
Section~4 contains a numerical comparison between the direct
numerical integration of the Feynman diagram and the results
of our analytic approximations.
Conclusions and future prospects are given
in section~4. 
In Appendix~A we collect some relevant information  
on one-loop two-point integrals with different masses and
present some formulae to handle the numerators. 
In Appendix~B we present a result for one of the 
next-to-leading terms of the expansion.

\section*{}
\begin{center}
{\bf 2. Constructing the expansion}
\end{center}


In Fig.~1a,b two different topologies of two-loop self-energy
diagrams are shown. The dimensionally-regularized scalar
Feynman integrals corresponding to these diagrams can be
written as\footnote{In many cases, we adopt the notation
used in the previous papers \cite{DT1,DST,BDST}.}
\be
\label{defJ}
J\left( \{\nu_i\} ; \{m_i\} ; k \right)
= \int \int 
\frac{\mbox{d}^n p \; \mbox{d}^n q}
{D_1^{\nu_1} D_2^{\nu_2} D_3^{\nu_3} D_4^{\nu_4} D_5^{\nu_5}} ,
\ee
where $n= 4-2\ep$ is the space-time dimension \cite{dimreg}, 
and $(D_i)^{\nu_i} \equiv (p_i^2 - m_i^2 + i0)^{\nu_i}$ are the powers
of the denominators (scalar propagators) corresponding 
to the internal lines of 
the diagrams in Fig.~1a,b. The momenta $p_i$ flowing through
the internal lines are constructed, in an obvious way, from
the external momentum $k$ and the loop momenta $p$ and $q$. 

We are mainly interested in the cases when all $\nu$'s are integer.
As mentioned in refs.~\cite{DT1,DST,BDST}, in this case it is
enough to consider the ``master'' diagram (Fig.~1a) only,
because (i) the diagrams with four and three internal lines 
can be obtained from Fig.~1a by shrinking (reducing to points) 
some of the internal lines (this corresponds to putting the 
corresponding $\nu$'s equal to zero)
and (ii) the diagram in Fig.~1b can be reduced to the 
diagrams with four denominators\footnote{For the
diagram in Fig.~1b, $p_1=p_4$. If, in addition, $m_1=m_4$, we get 
the integral with four propagators (with one the powers 
equal to  $\nu_1+\nu_4$). If $m_1 \neq m_4$
and both $\nu_1$ and $\nu_4$ are integer, one can use
standard partial fractioning to reduce the diagram in Fig.~1b
to a linear combination of the integrals with $\nu_1$
or $\nu_4$ equal to zero. All these integrals can be obtained
from the diagram in Fig.~1a by shrinking one of the lines.}.
Thus, in what follows we shall imply that the integral (\ref{defJ})
corresponds to the diagram in Fig.~1a.

In general, the diagram in Fig.~1a has two two-particle thresholds
(at $k^2=(m_1+m_4)^2$ and $k^2=(m_2+m_5)^2$) 
and two three-particle thresholds (at $k^2=(m_1+m_3+m_5)^2$
and $k^2=(m_2+m_3+m_4)^2$). 
If all these physical thresholds
correspond to non-zero (positive) values of $k^2$,
the small momentum expansion is nothing but a Taylor expansion
in $k^2$. In \cite{DT1} a general algorithm for calculating 
the coefficients of this expansion was developed.
Otherwise, if we have one (or more) physical threshold(s) at $k^2=0$,  
the corresponding ``zero-threshold'' expansion (as $n\to 4$) involves 
also the terms with $\ln(-k^2)$ (and even  $\ln^2(-k^2)$ when two
physical thresholds vanish).
To calculate analytically the coefficients of zero-threshold
expansion of the diagram of Fig.~1a, in \cite{BDST} 
general theorems on asymptotic expansions of Feynman diagrams 
\cite{as-ex} were applied.
In such a way, all different configurations of massless
thresholds have been considered. In particular, it has been verified 
that the expansion converges up to the first non-zero threshold.

In this paper, we study the behaviour of two-loop self-energy
diagrams which have small (but non-zero) physical thresholds. 
Namely, let us consider some of the masses corresponding to internal
lines of Fig.~1a to be large, while the other masses as well as the 
external momentum are small (and of the same order). To distinguish 
between the small and the large masses, we shall denote the latter 
with capital letters, $M_i$. By analogy with ref.~\cite{BDST}, 
four different small-threshold configurations exist, 
namely\footnote{2PT and 3PT mean two- and
three-particle thresholds, respectively.}:

Case 1: one small 2PT (e.g. masses $m_2$ and $m_5$ are small);

Case 2: two small 2PT's (the masses $m_1, m_2, m_4, m_5$ are small);

Case 3: one small 3PT (e.g. masses $m_2, m_3, m_4$ are small);

Case 4: one small 2PT and one small 3PT 
(e.g. the masses $m_2, m_3, m_4, m_5$ are small).

In addition, two special subcases of case~1, when one more
mass (not involved in the threshold) is small, should be considered 
separately:

Case 1a: case~1 with $m_3$ being small; 

Case 1b: case~1 with one of the masses involved in the 
second 2PT being small. 

In the present paper we restrict ourselves to cases~1, 1a, 1b and~2
where only small {\em two}-particle thresholds are involved 
(see a brief discussion of the cases 3 and 4 in the last section).

Before employing general results on
asymptotic expansions in momenta and masses \cite{as-ex}, we need to
introduce some notation. 
We shall need Taylor operators ${\cal T}_{k}$ and ${\cal T}_{m}$
expanding the denominators in the integrand 
(in the momentum or the mass, respectively) as
\be
\label{Taylor1}
{\cal T}_{k} \; \frac{1}{{[(k-p)^2 - m^2]}^{\nu}}
= \sum_{j=0}^{\infty} \frac{(\nu)_j}{j !} \;
    \frac{(2 (kp) - k^2)^j}{[p^2 - m^2]^{\nu+j}}  \; ,
\ee
\be
\label{Taylor2}
{\cal T}_{m} \; \frac{1}{[p^2-m^2]^{\nu}}
= \frac{1}{(p^2)^{\nu}} \;
\sum_{j=0}^{\infty} \frac{(\nu)_j}{j !}\; 
\left(\frac{m^2}{p^2}\right)^j ,
\ee
where
\be
\label{poch}
(\nu)_j \equiv \frac{\Gamma(\nu+j)}{\Gamma(\nu)}
\ee
is the Pochhammer symbol. The ``combined'' operators ${\cal T}$
expanding in some momenta and masses at the same time can
be constructed as products of the above operators (\ref{Taylor1})
and (\ref{Taylor2}). It is assumed that the Taylor operators are
applied to the integrand before the loop integration is performed.
After using eqs.~(\ref{Taylor1})--(\ref{Taylor2}), we should collect 
together all terms carrying the same {\em total} power of the small 
masses and momenta we expand in.

Let $\Gamma$ be the original
graph (corresponding in our case to Fig.~1a), subgraphs of $\Gamma$
are denoted as $\gamma$, and the corresponding ``reduced graph''
$\Gamma/\gamma$ is obtained from $\Gamma$ by shrinking the subgraph
$\gamma$ to a point. Furthermore, $J_{\gamma}$ is the
dimensionally-regularized Feynman integral with the denominators 
corresponding to a graph $\gamma$. In particular, $J_{\Gamma}$ 
corresponds to the integral (\ref{defJ}) itself.
Then, for our case, the general theorem yields
\be
\label{theorem}
J_{\Gamma}  \begin{array}{c} \frac{}{}  \\
                    {\mbox{\Huge$\sim$}} \\ {}_{k, \; m_i \to 0}
            \end{array}
\sum_{\gamma} J_{\Gamma / \gamma}
\circ {\cal T}_{k, \; m_{i}, \; q_{i}} \; J_{\gamma} ,
\ee
where the sum goes over all the subgraphs $\gamma$ which
(i) contain all the lines with the large masses $M_i$,
and (ii) are one-particle irreducible with respect to the
light lines (with the masses $m_i$).
The Taylor operator ${\cal T}_{k, \; m_{i}, \; q_{i}}$  
(see eqs.~(\ref{Taylor1})--(\ref{Taylor2}) )
expands the integrand of $J_{\gamma}$ in small masses $m_i$, 
the external momentum $k$ and the loop momenta $q_i$ 
which are ``external'' for a given subgraph $\gamma$.
The symbol ``$\circ$'' means that the polynomial in $q_i$,
which appears as a result of applying ${\cal T}$ to $J_{\gamma}$,
should be inserted in the numerator of the integrand
of $J_{\Gamma / \gamma}$. 

The definition of the class of subgraphs $\gamma$ involved in
the expansion (\ref{theorem}) is exactly the same as in 
the zero-threshold expansion
\cite{BDST}, just because zero masses are always small.
A distinction is that some tadpole-like contributions (vanishing
in zero-threshold limit) should be taken into account for
small non-zero masses (see below).


Let us consider which subgraphs $\gamma$ contribute to the sum
(\ref{theorem}) for the different small-threshold configurations.
To indicate which lines are included in the subgraph,
we shall use the numbering of lines given in Fig.~1a.
For example, the ``full'' graph $\Gamma\equiv \{12345\}$ includes
all five lines, $\{134\}$ corresponds to a subgraph where
the lines 2 and 5 are ``omitted'', etc. The subgraphs contributing
to the expansion are shown in Fig.~2, where bold and narrow
lines correspond to the propagators with large and small masses,
respectively. Dotted lines indicate the lines omitted in the
subgraphs $\gamma$ (as compared with the ``full'' graph $\Gamma$),
i.e. they correspond to the reduced graphs $\Gamma/\gamma$.

Comparing Fig.~2 with the four first lines of Fig.~3 presented in
\cite{BDST} (zero-threshold expansion), one can see that Fig.~2
contains some additional subgraphs for the cases 1a and 1b
(namely, the second and the fourth subgraphs for each of these cases).
These subgraphs did not contribute to the zero-threshold expansion
since the corresponding reduced graphs $\Gamma/\gamma$ yielded
massless tadpole-like integrals and therefore vanished
in dimensional regularization \cite{dimreg}.  
Here, we get tadpoles with small masses, which do {\em not}
vanish and must be taken into account\footnote{Appearance 
of extra subgraphs for the cases~1a and 1b does not
contradict the fact that all the cases~1, 1a and 1b correspond
to the same small-threshold configuration (one small 2PT).
We shall discuss details of this connection in section~3.}.  

{\em Case 1}.
Two subgraphs contribute to the asymptotic expansion: \\
(i) $\gamma=\Gamma \; \;  \Rightarrow$
\bea 
\int\int \mbox{d}^n p \; \mbox{d}^n q
\frac{1}{[(p-q)^2-M_3^2]^{\nu_3} [p^2-M_4^2]^{\nu_4} }
\hspace{70mm}
\nn \\  
\times {\cal T}_{k,m_2,m_5}
\frac{1}{[q^2-m_5^2]^{\nu_5}[(k-p)^2-M_1^2]^{\nu_1} 
[(k-q)^2-m_2^2]^{\nu_2}} ;
\label{c1-1}
\eea 
(ii) $\gamma = \{ 134 \} \; \; \Rightarrow$
\bea 
\int\int \mbox{d}^n p \; \mbox{d}^n q
\frac{1}{ [(k-q)^2-m_2^2]^{\nu_2}[q^2-m_5^2]^{\nu_5}}
\hspace{70mm}
\nn \\  
\times {\cal T}_{k,q}
\frac{1}{[(k-p)^2-M_1^2]^{\nu_1}[(p-q)^2-M_3^2]^{\nu_3}
[p^2-M_4^2]^{\nu_4}} . \label{c1-2}
\eea 

{\em Case 1a}.
Four subgraphs contribute to the asymptotic expansion: \\
(i) $\gamma=\Gamma \; \;  \Rightarrow$
\bea 
\int\int \mbox{d}^n p \; \mbox{d}^n q
\frac{1}{[p^2-M_4^2]^{\nu_4} }
\hspace{100mm}
\nn \\  
\times {\cal T}_{k,m_2,m_3,m_5}
\frac{1}{[(p-q)^2-m_3^2]^{\nu_3} [q^2-m_5^2]^{\nu_5} 
         [(k-p)^2-M_1^2]^{\nu_1} [(k-q)^2-m_2^2]^{\nu_2}} ; 
\label{c1a-1}
\eea 
(ii) $\gamma=\{ 1245 \} \; \;  \Rightarrow$
\bea 
\int\int \mbox{d}^n p \; \mbox{d}^n q
\frac{1}{[p^2-m_3^2]^{\nu_3} }
\hspace{100mm}
\nn \\  
\times {\cal T}_{k,p,m_2,m_5}
\frac{1}{[(p+q)^2-M_4^2]^{\nu_4} 
[q^2-m_5^2]^{\nu_5}[(k-p-q)^2-M_1^2]^{\nu_1} 
[(k-q)^2-m_2^2]^{\nu_2}} ; 
\label{c1a-2}
\eea 
(iii) $\gamma = \{ 134 \} \; \; \Rightarrow$
\bea 
\int\int \mbox{d}^n p \; \mbox{d}^n q
\frac{1}{ [(k-q)^2-m_2^2]^{\nu_2}[q^2-m_5^2]^{\nu_5}}
\hspace{70mm}
\nn \\  
\times {\cal T}_{k,q,m_3}
\frac{1}{[(k-p)^2-M_1^2]^{\nu_1}[(p-q)^2-m_3^2]^{\nu_3}
[p^2-M_4^2]^{\nu_4}} ; 
\label{c1a-3}
\eea 
(iv) $\gamma=\{ 14 \} \; \;  \Rightarrow$
\bea 
\int\int \mbox{d}^n p \; \mbox{d}^n q
\frac{1}{[p^2-m_3^2]^{\nu_3} [q^2-m_5^2]^{\nu_5} 
[(k-q)^2-m_2^2]^{\nu_2}}
\hspace{50mm}
\nn \\  
\times {\cal T}_{k,p,q}
\frac{1}{[(p+q)^2-M_4^2]^{\nu_4} [(k-p-q)^2-M_1^2]^{\nu_1} } ; 
\label{c1a-4}
\eea 

{\em Case 1b}.
Four subgraphs contribute: \\
(i) $\gamma=\Gamma \; \;  \Rightarrow$
\bea 
\int\int \mbox{d}^n p \; \mbox{d}^n q
\frac{1}{[(p-q)^2-M_3^2]^{\nu_3}}
\hspace{90mm}
\nn \\  
\times {\cal T}_{k,m_2,m_4,m_5}
\frac{1}{[p^2-m_4^2]^{\nu_4}  [q^2-m_5^2]^{\nu_5}
[(k-p)^2-M_1^2]^{\nu_1} [(k-q)^2-m_2^2]^{\nu_2}} ; 
\label{c1b-1}
\eea 
(ii) $\gamma=\{ 1235 \} \; \;  \Rightarrow$
\bea 
\int\int \mbox{d}^n p \; \mbox{d}^n q
\frac{1}{[p^2-m_4^2]^{\nu_4}}
\hspace{100mm}
\nn \\  
\times {\cal T}_{k,p,m_2,m_5}
\frac{1}{ [(p-q)^2-M_3^2]^{\nu_3} [q^2-m_5^2]^{\nu_5}
[(k-p)^2-M_1^2]^{\nu_1} [(k-q)^2-m_2^2]^{\nu_2}} ; 
\label{c1b-2}
\eea 
(iii) $\gamma = \{ 134 \} \; \; \Rightarrow$
\bea 
\int\int \mbox{d}^n p \; \mbox{d}^n q
\frac{1}{ [(k-q)^2-m_2^2]^{\nu_2}[q^2-m_5^2]^{\nu_5}}
\hspace{50mm}
\nn \\  
\times {\cal T}_{k,q,m_4}
\frac{1}{[(k-p)^2-M_1^2]^{\nu_1}[(p-q)^2-M_3^2]^{\nu_3}
[p^2-m_4^2]^{\nu_4}} ; 
\label{c1b-3}
\eea 
(iv) $\gamma = \{ 13 \} \; \; \Rightarrow$
\bea 
\int\int \mbox{d}^n p \; \mbox{d}^n q
\frac{1}{[p^2-m_4^2]^{\nu_4} [(k-q)^2-m_2^2]^{\nu_2}
[q^2-m_5^2]^{\nu_5}}
\hspace{40mm}
\nn \\  
\times {\cal T}_{k,p,q}
\frac{1}{[(k-p)^2-M_1^2]^{\nu_1}[(p-q)^2-M_3^2]^{\nu_3}} . 
\label{c1b-4}
\eea 

{\em Case 2}.
Four subgraphs contribute: \\
(i) $\gamma=\Gamma \; \; \Rightarrow$
\bea 
\int\int \mbox{d}^n p \; \mbox{d}^n q
\frac{1}{[(p-q)^2-M_3^2]^{\nu_3} }
\hspace{90mm}
\nn \\  
\times {\cal T}_{k,m_1,m_2,m_4,m_5}
\frac{1}{[p^2-m_4^2]^{\nu_4} [q^2-m_5^2]^{\nu_5}
[(k-p)^2-m_1^2]^{\nu_1} [(k-q)^2-m_2^2]^{\nu_2}} ; 
\label{c2-1}
\eea 
(ii) $\gamma = \{ 134 \} \; \; \Rightarrow$
\bea 
\int\int \mbox{d}^n p \; \mbox{d}^n q
\frac{1}{ [(k-q)^2-m_2^2]^{\nu_2}[q^2-m_5^2]^{\nu_5}}
\hspace{50mm}
\nn \\  
\times {\cal T}_{k,q,m_1,m_4}
\frac{1}{[(k-p)^2-m_1^2]^{\nu_1}[(p-q)^2-M_3^2]^{\nu_3}
[p^2-m_4^2]^{\nu_4}} ; \label{c2-2}
\eea 
(iii) $\gamma = \{ 235 \}$ contribution can be obtained from 
the previous one by the permutation 
$1 \leftrightarrow 2, 4 \leftrightarrow 5$; \\
(iv) $\gamma = \{ 3 \} \; \; \Rightarrow$
\be 
\int\int \mbox{d}^n p \; \mbox{d}^n q
\frac{1}{[(k\!-\!p)^2-m_1^2]^{\nu_1} [(k\!-\!q)^2-m_2^2]^{\nu_2}
         [p^2-m_4^2]^{\nu_4} [q^2-m_5^2]^{\nu_5}} 
\; {\cal T}_{p,q}
\frac{1}{[(p\!-\!q)^2-M_3^2]^{\nu_3}} . 
\label{c2-4}
\ee 

All the formulae~(\ref{c1-1})--(\ref{c2-4}), along with
eq.~(\ref{theorem}), can be used for any values of the
space-time dimension $n$ and the powers of propagators $\nu_i$.
Note that contributions from different subgraphs possess ultraviolet
and/or infrared poles in $\ep$. 
Cancellation of these poles in the
sum (when a finite integral is considered) turns out to be a very 
strong check of the calculational procedure\footnote{See also a
discussion in ref.~\cite{BDST}, p.~541.}.

Studying eqs.~(\ref{c1-1})--(\ref{c2-4}) shows that,
after partial fractioning is performed wherever necessary,
the following types of contributions can occur in the expressions
for the coefficients of the small-threshold expansion
(in situations with two-particle small thresholds): \\
(a) two-loop vacuum diagrams with two (or one) large-mass lines 
    and one (or two) massless lines;\\
(b) products of a one-loop massive diagram (with small masses and  
    external momentum $k$) and a one-loop massive tadpole;\\
(c) products of two one-loop massive diagrams
    with external momentum $k$.

The contributions of type (a) are discussed in detail in
Appendix~A of \cite{BDST}. The one-loop two-point functions
involving small masses of internal lines (occurring in the
contributions of types (b) and (c) ) should be calculated
exactly, and they are ``responsible'' for describing the
two-particle threshold irregularities. Some technical issues 
related to the calculation of these contributions are
collected in Appendix~A.



\section*{}
\begin{center}
{\bf 3. Analytic results}
\end{center}

To get analytic results for the terms of the expansion 
describing the small-threshold behaviour, we have used 
the {\sf REDUCE} system for analytical calculations \cite{Reduce}.
The constructed algorithm works for any (integer) powers 
of the propagators and can be applied for both convergent
and divergent diagrams.

As an example, in this section we consider
the diagram in Fig.~1a with unit powers of propagators,
the so-called ``master'' diagram. 
The corresponding scalar integral is involved in many interesting 
physical applications.
If all five $\nu_i$ are equal
to one the corresponding diagram 
is finite\footnote{To be precise, there is a very special 
case when this ``master'' integral becomes infrared-divergent: 
when there is a two-particle {\em zero}-threshold and, in addition,
$k^2=0$. The exact result for this case can be easily obtained
by a direct calculation.}  
as $n \to 4$, and we shall calculate the corresponding results
in four dimensions.
The algorithm makes it also possible to consider
higher terms of the expansion in $\ep = {\textstyle{1\over2}}(4-n)$, 
or even results for arbitrary $n$. 

Let us write the ``master'' integral as\footnote{In eq.~(\ref{J-exp}),
$m_i$ may correspond to either small or large masses.
In the expressions for the $S_j$ presented below, we
distinguish the small masses and the large masses by denoting
them as $m_j$ and $M_j$, respectively (as in section~2).} 
\be
\label{J-exp}
J(m_1,m_2,m_3,m_4,m_5; k)
\equiv J(1,1,1,1,1;m_1,m_2,m_3,m_4,m_5; k) =
- \pi^4 \sum_{j=0}^{\infty} S_j \; , 
\ee
where $S_j$ are the terms corresponding to our expansion.
For a given $j$, the term $S_j$ is a sum of all contributions
of the order $(k^2)^{j_0} \prod (m_i^2)^{j_i}$ (where the product
is taken over all lines with small masses) with
$j_0+ \sum j_i = j$. Since $k^2 J$ is dimensionless, so should be
$k^2 S_j$. 
Therefore, the $S_j$ should involve the corresponding
powers of the large masses in the denominators. 
The numerators of $S_j$ may also contain polynomials in the large masses,
logarithms of the ratios of the masses and some functions corresponding
to the contributions (a), (b), (c) described at the end of section 2. 
Note that (in the case when there are no zero thresholds) we should not
get the terms like $\ln(-k^2)$ producing a cut in the $k^2$ plane
starting from the origin.

The contributions of type (a) are expressed through the following
dimensionless function of two variables:
\be
\label{defH}
{\cal{H}}(M_1^2, M_2^2) = 2\Li{2}{1- \frac{M_1^2}{M_2^2}}
   + {\textstyle{1\over2}} \ln^2\left(\frac{M_1^2}{M_2^2}\right).
\ee
It is  antisymmetric
in its arguments and therefore vanishes when $M_1^2=M_2^2$.
This function is connected with the
finite part of the two-loop vacuum integral with two general masses 
and one zero mass (for details of this definition, see Appendix~A 
of \cite{DST}). In the contributions considered in this paper
the ${\cal{H}}$ function depends on the large masses only.

The contributions of types (b) and (c) are expressed through the
function $\tau(m_1, m_2; k^2)$
which is related to the finite part 
of the one-loop self-energy with general masses 
(see eq.~(\ref{def_tau})).
It is well-known that the $\tau$ function can be expressed in terms
of logarithms and square roots (see e.g. in \cite{'tHV'79}).
By using the notation
\be
\label{Delta}
\Delta(m_1^2, m_2^2, k^2) \equiv 4 m_1^2 m_2^2 - (k^2-m_1^2-m_2^2)^2
\ee
for the ``triangle'' K\"allen function\footnote{The function
(\ref{Delta}) is totally symmetric with respect to all its arguments.
It vanishes at the threshold, $k^2=(m_1+m_2)^2$, and at the 
pseudo-threshold, $k^2=(m_1-m_2)^2$.}, 
the result for the
$\tau$ function can be presented as
\bea
\label{tau}
\tau(m_1, m_2; k^2) 
= \frac{1}{2 k^2} \left\{ \sqrt{-\Delta} \;
\ln\frac{k^2-m_1^2-m_2^2-\sqrt{-\Delta}}
        {k^2-m_1^2-m_2^2+\sqrt{-\Delta}}
+ (m_1^2 - m_2^2) \ln\frac{m_2^2}{m_1^2}
\right.
\nonumber \\
\left.
+ \mbox{i} \pi \sqrt{-\Delta} \; 
\theta\left(k^2-(m_1+m_2)^2\right) \frac{}{} \right\} ,
\eea
where $\Delta\equiv \Delta(m_1^2, m_2^2, k^2)$.
In our contributions, the $\tau$ function depends on the small masses
only.
In fact, it contains the main information
about the small-threshold behaviour at two-particle thresholds.
The $\theta$ term in the braces yields an imaginary part in the
region beyond the physical threshold (i.e. for $k^2 > (m_1+m_2)^2$).
This is exactly the point where the cut in the complex $k^2$
plane starts.
Further properties of one-loop self-energy diagrams and the
$\tau$ function are discussed in Appendix~A.

Now, let us present explicit results for some lowest terms of
the expansion (\ref{J-exp}).  

{\em Case 1}. For the case of general masses, the $S_0$ contribution
to eq.~(\ref{J-exp}) is given by
\bea
\label{S0-case1}
S_0 =
\frac{1}{4 (M_1^2-M_3^2) (M_4^2-M_3^2) (M_1^2-M_4^2)}
\hspace{78mm}
\nonumber \\
\times\!
\left\{ -4 \!\left( \!\tau(m_2,m_5;k^2)
        \!+\!{\textstyle{1\over2}}
        \ln\frac{M_1^2 M_4^2}{m_2^2 m_5^2} \!+\!2 \!\right)\!
\left[ M_1^2 (M_4^2\!-\!M_3^2) \ln\frac{M_1^2}{M_3^2}
      -\! M_4^2 (M_1^2\!-\!M_3^2) \ln\frac{M_4^2}{M_3^2} \right]
\right.
\nonumber \\
+(M_1^2-M_3^2) (M_4^2-M_3^2) \ln\frac{M_1^2}{M_4^2}
    \left( \ln\frac{M_1^2}{M_3^2}+\ln\frac{M_4^2}{M_3^2} \right)
-2 M_3^2 (M_1^2-M_4^2) \ln\frac{M_1^2}{M_3^2} \ln\frac{M_4^2}{M_3^2}
\nonumber \\
\left.
\frac{}{}
+2 (M_1^2+M_3^2) (M_4^2-M_3^2) {\cal{H}}(M_1^2,M_3^2)
-2 (M_4^2+M_3^2) (M_1^2-M_3^2) {\cal{H}}(M_4^2,M_3^2)
\right\} \; ,
\hspace{5mm}
\eea
We note that this result is very similar to the corresponding 
contribution for the zero-threshold case, $C_0$ 
(see eq.~(21) of ref.~\cite{BDST}).
Indeed, in the massless limit we get (cf. eq.~(\ref{m1,m2->0}))
\be
\label{massless_tau}
\left. \frac{}{} \tau(m_1,m_2;k^2) \right|_{m_1, m_2 \to 0}
= {\textstyle{1\over2}} \left( 
\ln\left(-\frac{m_1^2}{k^2}\right)+
\ln\left(-\frac{m_2^2}{k^2}\right) 
\right) .
\ee
Therefore, the massless limit of the brackets in (\ref{S0-case1}) 
containing the $\tau$ function yields the same
combination involving $\ln(-k^2)$ as in eq.~(21) of \cite{BDST}, 
while the remaining terms coincide.
The next term, $S_1$, involves the $k^2$, $m_2^2$ and $m_5^2$
contributions 
(we present the corresponding result in Appendix~B). Note that in
this case the coefficient of $k^2$ gives 
(in the massless limit, i.e.\ using eq.~(\ref{massless_tau}))
the corresponding zero-threshold coefficient $C_1$ (see eq.~(22) of
\cite{BDST}), while the $m_2^2$ and $m_5^2$ contributions have 
no analogy with the zero-threshold case.

We have also obtained higher terms of the expansion (\ref{J-exp})
with five different masses ($S_2$ and $S_3$), but they are more 
cumbersome, and we do not present them here.

When some large masses are equal, one can use this fact
from the very beginning, when calculating contributions of the 
corresponding subgraphs. Another possibility is to consider, with
due care (since both denominators and numerators vanish in this limit),
the corresponding limit of the expressions with different masses.
On one hand, the second option is a good way to check the consistency
of the obtained results. On the other hand, the first way (i.e. having  
fewer different masses from the very beginning) simplifies
the calculation and therefore makes it possible to obtain expressions
for higher terms. 

For example, we get
the  following expressions for the first two coefficients 
corresponding to case 1 with $M_1=M_4 \equiv M$: 
\bea
\label{S0-case1_4m}
S_0 = -\frac{1}{(M^2-M_3^2)^2}
 \left\{ \left( \tau(m_2,m_5;k^2)
        +{\textstyle{1\over2}} \ln\frac{M^4}{m_2^2 m_5^2} \right)
    \left( M_3^2 \ln\frac{M_3^2}{M^2}+M^2-M_3^2 \right)
\right.
\nonumber \\
\left.
+ M_3^2 \left( {\cal{H}}(M^2,M_3^2)
               +{\textstyle{1\over2}} \ln^2\frac{M_3^2}{M^2} \right) 
   +2 (M^2-M_3^2) 
\right\} ,
\eea
\bea
\label{S1-case1_4m}
S_1 = -\frac{1}{12 M^2 (M^2-M_3^2)^4}
\hspace{105mm}
\nonumber \\
\times\!
\left\{ \left( \tau(m_2,m_5;k^2)
        \!+\! {\textstyle{1\over2}} \ln\frac{M^4}{m_2^2 m_5^2} \right)
\! \left( 6 M^2 M_3^2 \ln\frac{M^2}{M_3^2} 
      \left( k^2 M_3^2-(m_2^2+m_5^2) (2 M^2+M_3^2) \right)
\right.
\right.
\nonumber \\
\left. \frac{}{}
+(M^2\!-\!M_3^2) \left( k^2 (M^4\!-\!5 M^2 M_3^2\!-\!2 M_3^4)
        \!+\!3 (m_2^2\!+\!m_5^2) M^2 (M^2\!+\!5 M_3^2) \right) \right)
\nonumber \\
-6 M^2 M_3^2 \left({\cal{H}}(M^2,M_3^2)
               +{\textstyle{1\over2}} \ln^2\frac{M_3^2}{M^2} \right)
\left( k^2 M_3^2-2 (m_2^2+m_5^2) (2 M^2+M_3^2) \right)
\hspace{13mm}
\nonumber \\ 
+3 M^2 \left( \!m_2^2 \ln\frac{m_2^2}{M^2}
             \!+\!m_5^2 \ln\frac{m_5^2}{M^2} \!\right)\!
\left(\! 2 M_3^2 (2 M^2\!+\!M_3^2) \ln\frac{M^2}{M_3^2}
             \!-\!(M^2\!-\!M_3^2) (M^2\!+\!5 M_3^2) \! \right)\!
\nonumber \\ 
+2 M_3^2 \ln\frac{M^2}{M_3^2} 
\left( k^2 (5 M^4-M_3^4)+12 (m_2^2+m_5^2) M^2 M_3^2 \right)
\hspace{47mm}
\nonumber \\ 
\left.
+(M^2-M_3^2) \left( 6 (m_2^2+m_5^2) M^2 (M^2+7 M_3^2)
              -k^2 (M^4+17 M^2 M_3^2+2 M_3^4) \right) 
\frac{}{} \right\} .
\hspace{6mm}
\eea
For this case, we have also obtained the $S_2, \; S_3$ and $S_4$ 
terms. 

If all large masses are equal (i.e. $M_1=M_3=M_4\equiv M$), 
the expressions become very simple:
\be
S_0 =
- \frac{1}{2 M^2}
\left( \tau(m_2,m_5;k^2)  
       + {\textstyle{1\over2}} \; \ln\frac{M^4}{m_2^2 m_5^2} 
       + 3 \right) \; ,
\ee
\bea
S_1 =
-\frac{1}{144 M^4}
\left\{
6 (k^2 + m_2^2 + m_5^2)
 \left( \tau(m_2,m_5;k^2)
       + {\textstyle{1\over2}}\; \ln \frac{M^4}{m_2^2 m_5^2} \right)
\right.
\hspace{11mm} 
\nonumber \\ 
\left.
+ 6 \left( m_2^2 \ln\frac{M^2}{m_2^2} + m_5^2 \ln\frac{M^2}{m_5^2} \right)
+ 9 k^2 + 26 (m_2^2+m_5^2)
\right\} .
\eea
For this case, we have obtained the terms up to (and including) $S_6$.

{\em Case 1a.} For general masses, the lowest case~1a contributions
to (\ref{J-exp}) are  
\be
\label{S0-case1a}
S_0 = -\frac{1}{M_1^2-M_4^2} \ln\frac{M_1^2}{M_4^2}
\left( \tau(m_2,m_5;k^2)
+{\textstyle{1\over2}} \ln\frac{M_1^2 M_4^2}{m_2^2 m_5^2}+2
\right) ,
\ee
\bea
\label{S1-case1a}
S_1 = -\frac{1}{4 M_1^2 M_4^2 (M_1^2-M_4^2)^3}
 \left\{ 
2 \left( \tau(m_2,m_5;k^2)
         +{\textstyle{1\over2}} 
\ln\frac{M_1^2 M_4^2}{m_2^2 m_5^2}+2 \right)
\right.
\hspace{33mm}
\nonumber \\
\times \left[ M_1^2 M_4^2 
\left( 
\left( (M_1^2+M_4^2) k^2+(M_1^2-M_4^2) (m_2^2-m_5^2) \right)
\ln\frac{M_1^2}{M_4^2} 
-2 (M_1^2-M_4^2) k^2 \right)
\right.
\nonumber \\
\left.
+ (M_1^2-M_4^2)^2 
\left( 2 m_3^2 \left( M_1^2 \ln\frac{m_3^2}{M_4^2}
                         -M_4^2 \ln\frac{m_3^2}{M_1^2} \right)
- M_4^2 m_2^2+M_1^2 m_5^2 \right)
\right]
\nonumber \\ 
+ M_1^2 M_4^2 (M_1^2-M_4^2) \ln\frac{M_1^2}{M_4^2}
\left( m_2^2 \ln\frac{m_2^4}{M_1^2 M_4^2}
       -m_5^2 \ln\frac{m_5^4}{M_1^2 M_4^2}-2 m_2^2+2 m_5^2 \right)
\hspace{10mm}
\nonumber \\
-(M_1^2-M_4^2)^2
\left[ M_1^2 m_2^2 \ln\frac{m_2^4}{M_1^2 M_4^2}
       -M_4^2 m_5^2 \ln\frac{m_5^4}{M_1^2 M_4^2}           
       + (m_2^2+m_5^2) (M_1^2+M_4^2) \ln\frac{M_1^2}{M_4^2}  
\right.
\nonumber \\
+2 m_3^2 \ln\frac{M_1^2}{M_4^2}
\left( M_1^2 \ln\frac{m_3^2}{M_4^2}
      +M_4^2 \ln\frac{m_3^2}{M_1^2} \right)
+4 m_3^2 
\left( M_1^2 \ln\frac{m_3^2}{M_4^2}
      -M_4^2 \ln\frac{m_3^2}{M_1^2} \right)
\nonumber \\
\left.  \left.
+(M_1^2-M_4^2) 
\left( k^2-m_2^2-m_5^2
         +4 m_3^2 ({\textstyle{1\over3}} \pi^2-1) \right)
        -2 M_1^2 m_2^2+2 M_4^2 m_5^2 
\frac{}{} \right] 
\right\} .
\eea
For the general case~1a, we have also obtained the terms up to
(and including) $S_5$.

If $M_1=M_4\equiv M$, the corresponding results are
\be
S_0 =
- \frac{1}{M^2} 
\left( \tau(m_2,m_5;k^2) + {\textstyle{1\over2}} \;
\ln\frac{M^4}{m_2^2 m_5^2} + 2 \right) \; ,
\ee
\bea
S_1 = -\frac{1}{12 M^4}
\left\{
\left( \tau(m_2,m_5;k^2) 
       + {\textstyle{1\over2}}\ln\frac{M^4}{m_2^2 m_5^2} \right)\!
\left( k^2+\!3 (m_2^2\!+\!m_5^2\!+\!4m_3^2) 
       + 12 m_3^2 \ln \frac{m_3^2}{M^2}  \right)
\right. 
\nonumber  \\ 
\left.
- 3 \left( m_2^2 \ln   \frac{m_2^2}{M^2}
+m_5^2 \ln   \frac{m_5^2}{M^2}   \right)
- k^2 + 6 (m_2^2+m_5^2+4 m_3^2) - 4 \pi^2 m_3^2 
\right\} .
\eea

{\em Case 1b.} The results for the lowest terms, $S_0$ and $S_1$, 
are given by
\bea
\label{S0-case1b}
S_0 = -\frac{1}{2 M_1^2 (M_1^2-M_3^2)}
\left\{ 2 M_1^2 \ln\frac{M_1^2}{M_3^2}
\left( \tau(m_2,m_5;k^2)
  +{\textstyle{1\over2}} \ln\frac{M_3^4}{m_2^2 m_5^2}+2 \right)
\right.
\hspace{24mm}
\nonumber \\
\left.
-(M_1^2+M_3^2) 
\left( {\cal{H}}(M_1^2,M_3^2)
       -{\textstyle{1\over2}} \ln^2\frac{M_1^2}{M_3^2} \right)
-{\textstyle{1\over3}} \pi^2 (M_1^2-M_3^2) \right\} ,
\eea
\bea
\label{S1-case1b}
S_1\! && =\frac{1}{4 M_1^4 M_3^2 (M_1^2-M_3^2)^3}
\left\{ -2 M_1^2 
\left( \tau(m_2,m_5;k^2)
      +{\textstyle{1\over2}} \ln\frac{M_3^4}{m_2^2 m_5^2}+2 \right)
\right.
\nonumber \\
&&\times    
\left[ M_1^2 M_3^2 
\left( \left( (M_1^2-M_3^2) (k^2-m_5^2)+(M_1^2+M_3^2) m_2^2 \right) 
            \ln\frac{M_1^2}{M_3^2}
 -2 (M_1^2-M_3^2) m_2^2 \right)
\right.
\nonumber \\
&&
\left.
{\hspace*{20mm}}
+(M_1^2-M_3^2)^2 
\left( 2 m_4^2 \left( M_1^2 \ln\frac{m_4^2}{M_3^2}
                     -M_3^2 \ln\frac{m_4^2}{M_1^2} \right)
                    -M_3^2 k^2+M_1^2 m_5^2 \right) 
\right]
\nonumber \\ 
&& +M_3^2 \left( {\cal{H}}(M_1^2,M_3^2)
             -{\textstyle{1\over2}} \ln^2\frac{M_1^2}{M_3^2} \right) 
\left[ (M_1^2-M_3^2)^2 (M_1^2+M_3^2) (k^2+2 m_4^2)
\right.
\nonumber \\
&&
\left.
{\hspace*{45mm}}
+2 M_1^2 M_3^2 (M_1^2-M_3^2) k^2 +4 M_1^4 M_3^2 (m_2^2+m_5^2) \right]
\nonumber \\  
&& + (M_1^2\!-\!M_3^2) k^2 
\left( 2 M_1^2 M_3^2 (M_1^2\!+\!M_3^2) \ln\frac{M_1^2}{M_3^2}
       +M_1^4 (M_1^2\!-\!3 M_3^2)
       +{\textstyle{1\over3}} \pi^2 M_3^2 (M_1^2\!-\!M_3^2)^2 \right)
\nonumber \\
&& +4 (M_1^2-M_3^2)^3 m_4^2 
\left( M_1^2 \ln\frac{m_4^2}{M_3^2}-M_1^2
       +{\textstyle{1\over6}} \pi^2 (2 M_1^2+M_3^2) \right)
\nonumber \\
&& +2 M_1^4 \left( 
-(M_1^2-M_3^2) m_2^2 \ln\frac{m_2^2}{M_3^2}
\left( M_3^2 \ln\frac{M_1^2}{M_3^2}-M_1^2+M_3^2 \right)
\right.
\nonumber \\
&&
\left.
{\hspace{20mm}}
+M_3^2 m_5^2 \ln\frac{m_5^2}{M_3^2}
\left( (M_1^2+M_3^2) \ln\frac{M_1^2}{M_3^2}-2 M_1^2+2 M_3^2 \right)
\right)
\nonumber \\
&&
+2 M_1^4 M_3^2 \ln\frac{M_1^2}{M_3^2}
\left( (2 M_1^2+M_3^2) m_2^2+M_3^2 (k^2+m_5^2) \right)
\nonumber \\
&&
\left. 
+M_1^4 (M_1^2-M_3^2) 
\left( (5 M_3^2- 3 M_1^2) m_2^2+(7 M_3^2-M_1^2) m_5^2 \right) 
\frac{}{} \right\} .
\eea
For the general case~1b, we have also obtained the $S_2$, $S_3$ 
and $S_4$ terms.

If $M_1=M_3\equiv M$, the results for $S_0$ and $S_1$ are
\be
S_0 =
- \frac{1}{M^2}
\left( \tau(m_2,m_5;k^2)  
      + {\textstyle{1\over2}}\;\ln\frac{M^4}{m_2^2 m_5^2}
+ {\textstyle{1\over6}} \pi^2 + 4 \right) \; ,
\ee
\bea
S_1 \!=
-\frac{1}{12 M^4}
\left\{
 \left( \tau(m_2,m_5;k^2)
       \!+\! {\textstyle{1\over2}} \ln\frac{M^4}{m_2^2 m_5^2} 
       \!+\! 1 \right)
\left[ 3k^2 \!+\! m_2^2 \!+\! 3 m_5^2
      \!+\! 12 m_4^2 \left( \ln\frac{m_4^2}{M^2} \!+ \!1 \right) \right]
\right.
\nonumber \\
\left.
- 3 m_2^2 \ln\frac{m_2^2}{M^2} - m_5^2 \ln\frac{m_5^2}{M^2}
- k^2 (\pi^2\!-\!10) - 6 m_4^2 (\pi^2\!-\!8) 
+ {\textstyle{11\over3}}(m_2^2\!+\!m_5^2) \right\} .
\eea

Before proceeding to the case~2, let us discuss a connection
between the results for the cases~1 and 1a,b.
All of them correspond to the same small-threshold configuration
(one small two-particle threshold), and one could expect that
there should be a transition between explicit expressions
for the terms of the expansion (\ref{J-exp}) for the case~1 and 
those for the cases~1a and 1b. In the zero-threshold case
\cite{BDST}, what we needed was just to put one more mass
(in the coefficients $C_j$ for the case~1) to be zero, and we arrived
at the results for the corresponding case~1a and case~1b coefficients.
When we deal with small (but non-zero) thresholds, the situation is
more tricky. We need to consider one more mass to be small,
but now the contributions proportional to this small mass will go to
the higher $S_j$. In addition, we need to expand the denominators
involving this small mass as well as the corresponding ${\cal{H}}$
function(s) involving a small and a large mass. 
Expansion of the denominators is trivial, while the ${\cal{H}}$
function can
be expanded using the following formula: 
\be
{\cal{H}}(M^2,m^2) = 
2 \sum_{l=1}^{\infty} \frac{1}{l^2} \left(\frac{m^2}{M^2}\right)^l
- 2 \ln\frac{m^2}{M^2} 
\sum_{l=1}^{\infty} \frac{1}{l} \left(\frac{m^2}{M^2}\right)^l
-{\textstyle{1\over2}} \ln^2\frac{m^2}{M^2} 
-{\textstyle{1\over3}} \pi^2 .
\ee    

So, if one wants to show the correspondence between the case~1
and the cases~1a,b , one should not consider the $S_j$
themselves but to collect the terms of required order in the 
whole sum (\ref{J-exp}) instead. In such a way, we have checked that
all available terms for the case~1 produce the corresponding
terms for the cases~1a and 1b. On one hand, the described
additional ``subexpansion'' in a small mass corresponds, in some 
sense, to additional subgraphs appearing in the cases~1a and 1b 
as compared with the case~1, see Fig.~2. On the other hand,
if one is interested in cases~1a and 1b, it is better to consider
these cases from the very beginning rather than get results
from the case~1 (since more terms are available for the 
cases~1a and 1b).   

{\em Case 2.} The results for the $S_0$ and $S_1$ terms of
the expansion (\ref{J-exp}) are 
\be
\label{S0-case2}
S_0 \!=
-\frac{1}{M_3^2} \!  
\left\{ 
\left( \tau(m_1,m_4;k^2) 
       \!+\! {\textstyle{1\over2}} \ln\frac{M_3^4}{m_1^2 m_4^2} 
       \!+\! 1 \right) 
\left( \tau(m_2,m_5;k^2) 
       \!+\! {\textstyle{1\over2}} \ln\frac{M_3^4}{m_2^2 m_5^2} 
         + 1 \right)
       \!+\! {\textstyle{1\over3}}  \pi^2 \!-\! 1 \right\} ,
\ee
\bea
\label{S1-case2}
S_1 = \frac{1}{2 M_3^4}
  \left\{ (k^2\!-\!m_1^2\!-\!m_2^2\!-\!m_4^2\!-\!m_5^2)
\frac{}{}
\right.
\hspace{85.6mm}
\nonumber \\
\times  
\left( \tau(m_1,m_4;k^2) 
       \!+\! {\textstyle{1\over2}}  \ln\frac{M_3^4}{m_1^2 m_4^2} 
       \!-\! {\textstyle{1\over2}} \right) 
\left( \tau(m_2,m_5;k^2) 
       \!+\! {\textstyle{1\over2}}  \ln\frac{M_3^4}{m_2^2 m_5^2} 
       \!-\! {\textstyle{1\over2}} \right) 
\nonumber \\
+ \left( \tau(m_1,m_4;k^2) 
       + {\textstyle{1\over2}} \ln\frac{M_3^4}{m_1^2 m_4^2} 
       - {\textstyle{1\over2}} \right)
\left( m_2^2 \ln\frac{m_2^2}{M_3^2} + m_5^2 \ln\frac{m_5^2}{M_3^2}
                         - {\textstyle{3\over2}} (m_2^2+m_5^2) + k^2 \right)
\nonumber \\ 
+ \left( \tau(m_2,m_5;k^2) 
       + {\textstyle{1\over2}} \ln\frac{M_3^4}{m_2^2 m_5^2} 
       - {\textstyle{1\over2}} \right)
\left( m_1^2 \ln\frac{m_1^2}{M_3^2} + m_4^2 \ln\frac{m_4^2}{M_3^2}
                         - {\textstyle{3\over2}}  (m_1^2+m_4^2) + k^2 \right)
\nonumber \\ 
- \left( {\textstyle{2\over3}} \pi^2 - {\textstyle{9\over2}} \right) 
  \left( m_1^2+m_2^2+m_4^2+m_5^2- {\textstyle{1\over2}} k^2 \right)
\hspace{70mm}
\nonumber \\
\left.
+\frac{(m_1^2\!-\!m_4^2) (m_2^2\!-\!m_5^2)}{k^2}\!
 \left( \tau(m_1,m_4;k^2)\!-\!\tau(m_1,m_4;0) \right) \! 
 \left( \tau(m_2,m_5;k^2)\!-\!\tau(m_2,m_5;0) \right) \! \right\} \! ,
\eea
where the value of the $\tau$ function at $k^2=0$ is
\be
\label{tau_k^2=0}
\tau(m_1,m_2;0) = -1 - {\textstyle{1\over2}}\;
\frac{m_1^2+m_2^2}{m_1^2-m_2^2} \;
\ln\frac{m_1^2}{m_2^2} .
\ee
For the general case~2, we have obtained the terms of the expansion
(\ref{J-exp}) up to (and including) $S_6$.

An interesting feature of eq.~(\ref{S1-case2}) is the appearance
of $k^2$ in the denominator. 
For higher terms, higher powers of $k^2$ in the denominator
occur\footnote{The corresponding contributions
vanish when either $m_1^2=m_4^2$ or $m_2^2=m_5^2$.},
e.g. $(k^2)^2$ for $S_2$, etc. 
For general masses, the powers of $k^2$ in the denominator can
be cancelled by considering the Taylor expansion of the
$\tau$ function in $k^2$. The simplest example can be
seen in eq.~(\ref{S1-case2}).  

To conclude this section, we would like to note that we have
successfully compared the massless limit of the small-threshold expansion
with {\em all} coefficients from the four first columns
(corresponding to the cases~1,1a, 1b and 2, respectively)
of Table~1 presented on p.~545 of ref.~\cite{BDST}.



\section*{}
\begin{center}
{\bf 4. Numerical results}
\end{center}

Let us show how the small-threshold expansion can be applied to 
obtain approximate numerical results for self-energy
diagrams. We shall use as examples  a number of mass configurations
corresponding to diagrams occurring in the Standard Model.
We take the masses of the $W$  boson,
and of the charmed, bottom and top quarks to be\footnote{In this
section, we adopt ``standard'' notation for the masses of physical
particles: the quark masses are denoted with small $m$ whereas
the capital $M$ is used for the vector boson masses. To avoid
any confusion with the notation of sections~2 and 3, we shall 
explicitly state, for each concrete example, which masses are
considered to be small and large.}:
\be
 M_W = \mbox{80 GeV}, \;\;
 m_c = \mbox{1.5 GeV}, \;\;
 m_b = \mbox{5 GeV},  \;\;
 m_t = \mbox{174 GeV}.
\ee
In this section, we shall consider the following approximations 
to the ``master'' integral (\ref{J-exp}) (corresponding to Fig.~1a):
\be
\label{J(N)}
J^{(N)} = - \pi^4 \sum_{j=0}^{N} S_j  \, .
\ee

As in ref.~\cite{BDST}, the first example corresponds to a diagram 
containing a top-bottom loop to which two $W$ bosons are attached. 
This diagram contributes to the
self-energy of the photon and the $Z$ boson. The corresponding
scalar integral is
\be
\label{eq:WbtWbk}
  J(M_W,m_b,m_t,M_W,m_b;k) \, .
\ee
If we consider $m_b$ as a small mass, and $m_t$ and $M_W$ as
large masses\footnote{Note that in ref.~\cite{BDST} we have 
considered the situation
when $m_b=0$. The situation for non-zero $m_b$ was discussed
as a subject to be investigated in the future. This paper
solves the problem posed in \cite{BDST}.},
this diagram has one small two-particle threshold 
at $k^2=4m_b^2$ and therefore it
belongs to case~1.
In Fig.~3
the approximations $J^{(N)}$
defined by eq.~(\ref{J(N)})
are shown as curves, and, for comparison, values of $J$ obtained by
numerical integration \cite{Kreimer} are shown as crosses. 
The position of the lowest ``large'' threshold, at $k^2=4M_W^2$
in this example, is indicated by a vertical line.
One can see that the expansion
converges all the way up to the
first large  threshold at $k^2=25600\mbox{ GeV}^2$.


In fact, the behaviour for $4m_b^2 \ll k^2 < 4M_W^2$ is described 
as well as in \cite{BDST}. The main difference is in the region 
of small $k^2$ (in particular, around the $4m_b^2$ threshold)
which could not be described by the zero-threshold expansion 
\cite{BDST}. 
In the scale of Fig.~3, the details of the small-threshold behaviour 
cannot be seen
so that we use a different pair of plots in Fig.~4 
to illustrate the behaviour in the region of small momenta,  
including the threshold at $k^2=4m_b^2=100\mbox{ GeV}^2$.
One can see that even the zero order approximation $J^{(0)}$ is 
very good and that it is sufficient to take $J^{(1)}$ to get 
result with a high precision. We do not present other approximations 
(up to $N=4$) because they give a precision better than the numerical
program does (and one cannot distinguish them in the plot).
In our approximations, the irregularities at the small threshold 
are reproduced by the $\tau$ function (\ref{tau}).
In particular, the imaginary part is zero below the threshold
and behaves like a square root in the region above the threshold. 


In fact, one can get a better description
of the behaviour of the same diagram (\ref{eq:WbtWbk}) in the
region around the $k^2=4M_W^2$ threshold by considering
$M_W$ as a small mass. In this case, the only remaining
large mass is $m_t$.
Thus, it belongs to case~2, in our terminology. According to our 
results, the situation in the region of small momenta 
is more or less the same: the first two approximations happen to be 
good enough (but not so good as when the diagram is considered
as case~1). Our approximations
in the region up to the large three-particle threshold (at
$k^2=(m_t+M_W+m_b)^2$) are shown in 
Fig.~5.
One can see that our approximations 
describe the behaviour around the $4M_W^2$ threshold and,
moreover, they work even beyond
this second threshold\footnote{The convergence between the $4M_W^2$
and the $(m_t+M_W+m_b)^2$ thresholds is not so good since the
$M_W/m_t$ ratio is not very small.}.
Thus, just by treating in another way
the available massive parameters it is possible to characterize 
analytically the first diagram in a larger region of momenta.


As second example  we consider the integral
\be
\label{eq:bcWbc}
  J(m_b,m_c,M_W,m_b.m_c;k) \, 
\ee
which has two small two-particle thresholds, at $k^2=4m_c^2$
and $k^2=4m_b^2$, and therefore belongs to case~2.

The behaviour up to the first large threshold 
is rather similar to the previous examples so we do not present 
the corresponding plot.
The behaviour at small momenta including both small thresholds
can be seen in 
Fig.~6. 
A rather tricky behaviour in the region around
these two small thresholds is very well reproduced
by taking the lowest analytical approximations, $J^{(0)}$
(for the imaginary part) or $J^{(1)}$ (for the real part). 

%


All the examples illustrate that the small-threshold expansion  
provides approximations which perfectly describe the small-threshold
behaviour. Moreover, they
are at least as accurate as numerical integration in a large
part of the region of convergence and can be evaluated much faster.

\newpage

\section*{}
\begin{center}
{\bf 5. Conclusions}
\end{center}

In this paper we have studied the behaviour of 
two-loop self-energy diagrams in the situation when the external momentum 
and some of the masses are small with respect to the large masses.
All configurations with small two-particle thresholds have been considered.
By use of explicit formulae for the terms of asymptotic expansions 
in the large mass limit, we presented an analytic approach to
calculating these diagrams by keeping the first few terms 
of the expansion.

By taking some complicated cases (corresponding to diagrams with different 
masses occurring in the Standard Model) as examples, we
compared our results with those of a numerical integration program 
based on the algorithm of ref.~\cite{Kreimer} 
(see also in \cite{berends2}).
We have shown that our analytical approximations work very well in the 
region of small momenta that includes all the small thresholds. 
Moreover, the 
small-threshold expansion converges up to a region close to
the first large threshold. Unless $k^2$ is close to
the large threshold, only a few terms are needed to obtain accurate results.
This comparison can also be considered as a check of the
numerical program. 

Thus we have solved the problem of the threshold behaviour 
of two-loop two-point functions for the small two-particle thresholds. 
It is interesting that it was possible to do this just by using the
expansion in the large masses. The main idea was to avoid putting
any conditions on relative values of the external momentum squared
and small masses.
To our knowledge,
no efficient algorithms for describing the non-zero threshold behaviour
of two-loop diagrams with arbitrary masses were available 
so far\footnote{A possible exception may be related to an interesting
approach considered in ref.~\cite{Tbilisi} where a two-loop diagram
with one non-zero mass parameter was studied as an example. 
It is not clear, however, how efficiently the method of \cite{Tbilisi}
will work for 
two-loop diagrams with different non-zero masses, since the
coefficients of the expansion may require results for the threshold
values of two-loop integrals (which, for the general case, are
not known analytically).}.
A nice feature of the presented approach is that all two-particle 
threshold irregularities of two-loop diagrams are incorporated 
in the function corresponding to one-loop diagrams,
the $\tau$ function.

A natural extension of this work could be connected with cases~3 and 4 
(according to the classification given in Section~2), when three-particle 
small thresholds arise. 
To carry out a similar program for those cases we can apply the same 
technique of asymptotic expansion in large mass(es). In particular,
in eq.~(\ref{theorem}) one should sum over the subgraphs obeying
the same conditions as in the cases considered. 
For example, for case~3 (with small masses $m_2, m_3$ and $m_4$)
the following five subgraphs $\gamma$ contribute to the sum 
(\ref{theorem}): $\Gamma\equiv\{12345\}$, $\{1235\}$, $\{1345\}$,
$\{1245\}$ and $\{15\}$. For case~4 (when all the masses except
$M_1$ are small), four subgraphs contribute: $\Gamma$, $\{1245\}$,
$\{134\}$ and $\{1\}$. 
However, in these cases the small-threshold behaviour at three-particle
threshold is defined by the functions corresponding to the sunset
diagram (with three propagators) and the diagram with four propagators
(e.g. the integral (\ref{defJ}) with $\nu_1=0$). 
Unfortunately, sufficient {\em analytic} information about 
threshold behaviour of these 
diagrams is at the moment not available.

The algorithm presented in this paper can be also extended to the
three-point two-loop diagrams with different masses.


\vspace{6mm}

{\bf Acknowledgements.}
A.~D. and V.~S. would like to thank the Instituut-Lorentz, 
University of Leiden, for hospitality during the visit when the largest 
part of this work was done.
We are grateful to J.B.~Tausk for his help and advices which were very 
useful, especially in the numerical part.
A.~D. and V.~S. are grateful to K.G.~Chetyrkin for useful discussion.

\section*{}
\begin{center}
{\bf Appendix A. One-loop two-point integrals with masses}
\end{center}

In this appendix, we collect some relevant results for the integrals 
\be
\label{defJ1}
J(\nu_1, \nu_2; m_1, m_2) = 
\int \frac{\mbox{d}^n q}
          {\left[q^2-m_1^2\right]^{\nu_1} \; 
           \left[(k-q)^2-m_2^2\right]^{\nu_2}} .
\ee

Let us introduce dimensionless quantities
\be
x \equiv \frac{m_1^2}{k^2}, \hspace{5mm} y \equiv \frac{m_2^2}{k^2},
\ee
\be
\lambda(x,y) = \sqrt{(1-x-y)^2 - 4xy} .
\ee
It is easy to see that the K\"allen function $\Delta$, 
eq.~(\ref{Delta}), is related to $\lambda$ via
\be
\label{Delta2}
\Delta(m_1^2, m_2^2, k^2)
= 4 m_1^2 m_2^2 - (k^2-m_1^2-m_2^2)^2 
= - (k^2)^2 \; \lambda^2(x,y) .
\ee
Therefore, $\lambda$ also vanishes at the threshold, 
$k^2=(m_1+m_2)^2$, and at the pseudo-threshold, $k^2=(m_1-m_2)^2$.

For arbitrary $\nu_1, \nu_2$ and the space-time dimension $n$,
the result for the integral (\ref{defJ1}) can be written as 
(see in \cite{BD})
\bea
J(\nu_1, \nu_2; m_1, m_2) = 
\pi^{n/2} \mbox{i}^{1-n} (k^2)^{n/2-\nu_1-\nu_2}
\hspace{78mm}
\nonumber \\
\times \left\{  
\frac{\Gamma\left(\frac{n}{2}-\nu_1\right)
      \Gamma\left(\frac{n}{2}-\nu_2\right)
      \Gamma\left(\nu_1+\nu_2-\frac{n}{2}\right)}
     {\Gamma(\nu_1) \; \Gamma(\nu_2) \; \Gamma(n-\nu_1-\nu_2)} \;
\right.
\hspace{72mm}
\nonumber \\
\times
F_4 \left( \left. \nu_1+\nu_2-{\textstyle{n\over2}}, \nu_1+\nu_2-n+1;
         \nu_1-{\textstyle{n\over2}}+1, \nu_2-{\textstyle{n\over2}}+1
           \right| x, y \right)
\hspace{4mm}
\nonumber \\
+ (-x)^{n/2 - \nu_1} \;
\frac{\Gamma\left(\nu_1-\frac{n}{2}\right)}{\Gamma(\nu_1)} \;
F_4\left( \left. \nu_2, \nu_2-{\textstyle{n\over2}}+1;
         {\textstyle{n\over2}}-\nu_1+1, \nu_2-{\textstyle{n\over2}}+1
           \right| x, y \right) 
\hspace{4mm}
\nonumber \\
\left.
+ (-y)^{n/2 - \nu_2} \;
\frac{\Gamma\left(\nu_2-\frac{n}{2}\right)}{\Gamma(\nu_2)} \;
F_4\left( \left. \nu_1, \nu_1-{\textstyle{n\over2}}+1;
         {\nu_1-\textstyle{n\over2}}+1, {\textstyle{n\over2}}-\nu_2+1
           \right| x, y \right)
\right\} ,
\eea
where $F_4$ is the Appell hypergeometric function of two variables,
\be
\label{F4}
\left. F_4\left(a,b; c,d \right| x, y \right)
= \sum_{j=0}^{\infty} \sum_{l=0}^{\infty}
\frac{(a)_{j+l} \; (b)_{j+l}}{(c)_j \; (d)_l}
\; \frac{x^j \; y^l}{j! \; l!} ,
\ee
and $(a)_j \equiv \Gamma(a+j)/\Gamma(a)$ is the Pochhammer symbol.

For $\nu_1=\nu_2=1$, but keeping the space-time dimension 
$n\equiv 4-2\ep$
arbitrary, the result simplifies and can be written in terms
of the Gauss hypergeometric function $_2F_1$,
\bea
J(1,1; m_1, m_2) = \mbox{i} \; \pi^{2-\ep} \; (-k^2)^{-\ep} \;
\left\{ \frac{\Gamma^2(1-\ep) \Gamma(\ep)}{\Gamma(2-2\ep)} \; 
\lambda^{1-2\ep}
\right.
\hspace{50mm}
\nonumber \\
- \Gamma(-1+\ep) \; {\textstyle{1\over2}}(1+x-y-\lambda) 
(-x)^{-\ep} \;
_2F_1\left( 
\left. \begin{array}{c} {1, \ep} \\ {2-\ep} \end{array} \right|
\frac{(1+x-y-\lambda)^2}{4x} \right)
\hspace{4mm}
\nonumber \\
\left.
- \Gamma(-1+\ep) \; {\textstyle{1\over2}}(1-x+y-\lambda) 
(-y)^{-\ep} \;
_2F_1\left( \left. 
\begin{array}{c} {1, \ep} \\ {2-\ep} \end{array} \right|
\frac{(1-x+y-\lambda)^2}{4y} \right)
\right\} .
\eea

Expanding in $\ep={\textstyle{1\over2}}(4-n)$ we get 
\be
\label{def_tau}
J(1,1; m_1, m_2)= \mbox{i} \pi^{2-\ep} \; \Gamma(1+\ep) \;
\left\{ \frac{1}{\ep} + 2 
- \textstyle{1\over2} \left( \ln{m_1^2} + \ln{m_2^2} \right)
+ \tau(m_1, m_2; k^2) + {\cal{O}}(\ep) \right\} .
\ee
where the $\tau$ function can be defined via an integral 
representation 
\be
\tau(m_1, m_2; k^2) \equiv \tau(x,y) = - \Int_0^1 \mbox{d} \alpha \;
\ln\left( 
\frac{\alpha m_1^2 + (1-\alpha) m_2^2 - \alpha (1-\alpha) k^2}
     {\alpha (1-\alpha) m_1 m_2} 
\right) .
\ee
It is understood that $k^2 \leftrightarrow k^2 +\mbox{i} 0$.

For $k^2<(m_1-m_2)^2$ (including the Euclidean region, $k^2<0$), the 
result for the $\tau$ function can be written as
\be
\tau(x,y) = {\textstyle{1\over2}} \lambda \; 
\ln\frac{1-x-y-\lambda}{1-x-y+\lambda}
+ {\textstyle{1\over2}} (x-y) \ln\frac{y}{x} .
\ee
This is also valid all the way up to the threshold, $k^2=(m_1+m_2)^2$.
However, since $\lambda(x,y)$ is imaginary for
$(m_1-m_2)^2 < k^2 < (m_1+m_2)^2$, $\lambda^2 <0$, in this region 
one can also use another representation,
\be
\tau(x,y) = - \sqrt{-\lambda^2} \; 
\arccos\left( \frac{x+y-1}{2\sqrt{x y}} \right)
+ {\textstyle{1\over2}} (x-y) \ln\frac{y}{x} .
\ee
Beyond the threshold, at $k^2 > (m_1+m_2)^2$, $\lambda^2$ is again 
positive, but the $\tau$ function obtains an imaginary part,
\be
\tau(x,y) = \mbox{i} \pi \lambda 
+ {\textstyle{1\over2}} \lambda \; 
\ln\frac{1-x-y-\lambda}{1-x-y+\lambda}
+ {\textstyle{1\over2}} (x-y) \ln\frac{y}{x} .
\ee

Note that at $\lambda=0$ the expression for the $\tau$ function is 
\be
\left. \frac{}{} \tau(x,y)\right|_{\lambda=0}
= {\textstyle{1\over2}} (x-y) \ln\frac{y}{x} ,
\ee
whereas at $k^2=0$ we get the result (\ref{tau_k^2=0}).

When one or two masses $m_i$ vanish, the $\tau$ function develops a
logarithmic singularity. For example, taking the limit $m_2 \to 0$
yields the well-known result
\be
\label{m2->0}
\left. \frac{}{} \tau(m_1, m_2; k^2)\right|_{m_2 \to 0}
= {\textstyle{1\over2}} \ln\frac{m_2^2}{m_1^2}
+ \frac{m_1^2-k^2}{k^2} \; \ln\frac{m_1^2-k^2}{m_1^2} .
\ee
The term $\textstyle{1\over2} \ln m_2^2$ cancels the corresponding 
term in (\ref{def_tau}). The rest is real for $k^2<m_1^2$, i.e. 
below the threshold (in this case the threshold and the 
pseudo-threshold coincide). For $k^2>m_1^2$ (beyond the threshold), 
we should remember that $k^2 \leftrightarrow k^2+\mbox{i}0$ and 
substitute
$\ln(m_1^2-k^2) \leftrightarrow \ln(k^2-m_1^2) - \mbox{i}\pi$.

If both masses vanish, we get
\be
\label{m1,m2->0}
\left. \frac{}{} \tau(m_1,m_2; k^2)\right|_{m_1, m_2 \to 0}
= {\textstyle{1\over2}} \left( \ln\left(-\frac{m_1^2}{k^2}\right)
        + \ln\left(-\frac{m_2^2}{k^2}\right) \right) .
\ee
The terms $\textstyle{1\over2}(\ln m_1^2 + \ln m_2^2)$
cancel the corresponding terms in (\ref{def_tau}), while
the remaining $\ln(-k^2)$ should be understood as
$(\ln k^2 - \mbox{i} \pi)$ for the time-like values of the momentum.
In this limit, the results of the present paper correspond to the
results of zero-threshold expansion \cite{BDST}.

The integrals $J(\nu_1, \nu_2; m_1, m_2)$ with higher integer values
of $\nu_1$ or $\nu_2$ can be obtained by using the integration-by-parts
technique \cite{ibp}. The way is quite similar to one used 
in \cite{JPA}. First, we get a system of two equations,
\bea
-2\nu_1 m_1^2 J(\nu_1+1, \nu_2) 
+ \nu_2 (k^2-m_1^2-m_2^2) J(\nu_1, \nu_2+1)
\hspace{50mm}
\nonumber \\
= -(n-2\nu_1-\nu_2) J(\nu_1, \nu_2) + \nu_2 J(\nu_1-1,\nu_2+1),
\nonumber \\[2mm]
\nu_1 (k^2-m_1^2-m_2^2) J(\nu_1+1, \nu_2) 
-2\nu_2 m_2^2 J(\nu_1, \nu_2+1)
\hspace{50mm}
\nonumber \\
= -(n-\nu_1-2\nu_2) J(\nu_1, \nu_2) + \nu_1 J(\nu_1+1, \nu_2-1) .
\eea 
The determinant of the matrix composed of the coefficients
on the l.h.s. of this system is proportional to
$\Delta(m_1^2, m_2^2, k^2)$, eq.~(\ref{Delta2}).
Solving the system, we get
\bea
J(\nu_1+1, \nu_2) = \frac{1}{\nu_1 \Delta}
\left\{ \left( (k^2-m_1^2) (n-\nu_1-2\nu_2)+m_2^2 (n-3\nu_1) \right) 
J(\nu_1, \nu_2)
\right.
\hspace{20mm}
\nonumber \\
\left.
-2 \nu_2 m_2^2 J(\nu_1-1, \nu_2+1) 
- \nu_1 (k^2-m_1^2-m_2^2) J( \nu_1+1, \nu_2-1) \right\} ,
\eea
and an analogous result for $J(\nu_1, \nu_2+1)$.
By using these results, all the integrals with higher integer 
$\nu$'s can be expressed in terms of the integral 
$J(1,1; m_1, m_2)$ and the massive tadpoles. 

We also need some formulae for the integrals with numerators.
For the integrals (\ref{defJ1}) with one negative power of the
denominator, we get
\bea
J(\nu, -N; m_1, m_2) 
= \int \frac{\mbox{d}^n q}{\left[ q^2-m_1^2 \right]^{\nu}}
\; \left[ (k-q)^2 - m_2^2 \right]^N
\hspace{30mm}
\nonumber \\
= \sum_{j=0}^{[N/2]} \frac{N!}{j! (N-2j)! 
\left( \frac{n}{2} \right)_j} \;
(k^2)^j \; \int \mbox{d}^n q \; 
\frac{(q^2)^j \; (k^2+q^2-m_2^2)^{N-2j}}
     {\left[ q^2 - m_1^2 \right]^{\nu}} .
\eea

Another formula we would like to present is useful for dealing
with the two-loop integrals with the denominators corresponding
to a product of two one-loop integrals (\ref{defJ}), but when
the numerator is a power of the scalar product of two integration
momenta. We need such integrals to calculate the terms of the
expansion (\ref{J-exp}) for case~2. We shall present the
result valid for a more general case, when the integrand is 
a product of two arbitrary scalar functions. Let
\begin{equation}
K\left[ \mbox{something} \right]
= \int \int \mbox{d}^n p \; \mbox{d}^n q \;
\left\{ \mbox{something} \right\} \; 
F_1(k^2, p^2, (pk)) \; F_2(k^2, q^2, (qk)) .
\end{equation}
Then
\begin{eqnarray}
\label{Gegen}
K\left[ (pq)^N \right]
= \frac{N!}{2^N (k^2)^N} \sum_{\{j\}}^{}
\frac{{j_3}!}{j!} \;
\frac{\left(\frac{n}{2}\right)_{j_3}}
     {\left(\frac{n-2}{2}\right)_{j_3} \; (n-2)_{j_3} \;
        \left(\frac{n}{2}\right)_{j+j_3}} 
\hspace{50mm}
\nonumber \\
\times \;
K\left[ (k^2)^{N/2} (p^2)^{N/2} 
 C_{j_3}^{(n-2)/2}\left(\frac{(kp)}{\sqrt{k^2 p^2}}\right) \;
(k^2)^{N/2} (q^2)^{N/2} 
 C_{j_3}^{(n-2)/2}\left(\frac{(kq)}{\sqrt{k^2 q^2}}\right) 
\right] ,
\end{eqnarray}
where the sum goes over all $\{j,j_3\}$ such that $2j+j_3=N$
(so, in fact this is a one-fold finite sum) while\footnote{The 
symbol $[j/2]$ denotes the integer part of $j/2$.}
\be
\label{def-Gegen}
C_j^{\gamma}(x)=\sum_{l=0}^{[j/2]} 
\frac{(-1)^l (\gamma)_{j-l}}{l! (j-2l)!}
\; (2x)^{j-2l}
\ee
are Gegenbauer polynomials. The result (\ref{Gegen}) can be derived
using the formulae presented in ref.~\cite{ChKT}\footnote{It is 
interesting to note that eq.~(\ref{Gegen}) has a structure similar to
eq.~(43) of \cite{DT3}, with $N_1=N_2=N$ and $j_1=j_2=j$.}.
Using the definition (\ref{def-Gegen}), it is easy to see that
we do not get neither negative powers nor square roots of
$p^2$ and $q^2$ on the r.h.s. of eq.~(\ref{Gegen}).

\section*{}
\begin{center}
{\bf Appendix B. The $S_1$ term for the case~1 with general masses}
\end{center}

Here we present the result for the $S_1$ contribution to the
expansion (\ref{J-exp}) for the case~1 with different 
masses\footnote{The
lowest term, $S_0$, is given by eq.~(\ref{S0-case1}).}:
\bea
\label{S1-case1}
S_1 = -\frac{1}{4 (M_1^2-M_3^2)^3 (M_4^2-M_3^2)^3 (M_1^2-M_4^2)^3}
 \left\{ 
2 \left( \tau(m_2,m_5;k^2)
+{\textstyle{1\over2}} \ln\frac{M_1^2 M_4^2}{m_2^2 m_5^2}+2 \right)
\right.
\nonumber \\
\times
\left[ M_1^2 (M_4^2-M_3^2)^3 \ln\frac{M_1^2}{M_3^2}
\left( M_1^2 (M_1^2-M_3^2) (M_1^2-M_4^2) (k^2+m_2^2-m_5^2)
\right. \right.
\hspace{22mm}
\nonumber \\
\left.
+2 M_4^2 (M_1^2-M_3^2)^2 k^2+2 M_3^2 (M_1^2-M_4^2)^2 m_2^2 \right)
\nonumber \\
+M_4^2 (M_1^2-M_3^2)^3 \ln\frac{M_4^2}{M_3^2}
\left( M_4^2 (M_4^2-M_3^2) (M_1^2-M_4^2) (k^2-m_2^2+m_5^2)
\right.
\hspace{20mm}
\nonumber \\
\left.
-2 M_1^2 (M_4^2-M_3^2)^2 k^2-2 M_3^2 (M_1^2-M_4^2)^2 m_5^2 \right)
\nonumber \\
-(M_1^2-M_3^2) (M_4^2-M_3^2) (M_1^2-M_4^2)
\hspace{77mm}
\nonumber \\
\times
\left( (M_1^2-M_3^2) (M_4^2-M_3^2) 
\left( M_1^2 (M_4^2-M_3^2)+M_4^2 (M_1^2-M_3^2) \right) k^2
\right.
\hspace{8mm}
\nonumber \\
+(M_4^2-M_3^2) (M_1^2-M_4^2) 
\left( M_1^2 (M_4^2-M_3^2)-M_3^2 (M_1^2-M_4^2) \right) m_2^2
\hspace{5mm}
\nonumber \\
\left. \left.
-(M_1^2-M_3^2) (M_1^2-M_4^2) 
\left( M_3^2 (M_1^2-M_4^2)+M_4^2 (M_1^2-M_3^2) \right) m_5^2 \right) 
\right] 
\nonumber \\
-\left( {\cal{H}}(M_1^2,M_3^2)
       -{\textstyle{1\over2}} \ln^2\frac{M_1^2}{M_3^2} \right) 
      (M_4^2-M_3^2)^3
\left[ 
4 M_1^2 M_3^2 (M_1^2-M_4^2)^2 (m_2^2+m_5^2) 
\right.
\hspace{15mm}
\nonumber \\
\left. + (M_1^2-M_3^2) 
\left( (M_1^2+M_4^2) (M_1^4-M_3^4)+2 M_1^2 M_3^2 (M_1^2-M_4^2) 
\right) k^2
\right]
\nonumber \\
+\left( {\cal{H}}(M_4^2,M_3^2)
        -{\textstyle{1\over2}} \ln^2\frac{M_4^2}{M_3^2} \right) 
  (M_1^2-M_3^2)^3
\left[
4 M_4^2 M_3^2 (M_1^2-M_4^2)^2 (m_2^2+m_5^2) 
\right.
\hspace{15mm}
\nonumber \\
\left. 
+ (M_4^2-M_3^2) 
\left( (M_1^2+M_4^2) (M_4^4-M_3^4)-2 M_4^2 M_3^2 (M_1^2-M_4^2) 
\right) k^2
\right]
\nonumber \\ 
+2 (M_1^2-M_3^2) (M_1^2-M_4^2) m_2^2 \ln\frac{m_2^2}{M_3^2}
\hspace{86mm}
\nonumber \\
\times\!
\left[ (M_4^2\!-\!M_3^2)^3 M_1^4 \ln\frac{M_1^2}{M_3^2}
\!-\! (M_1^2\!-\!M_3^2)^2 
\left( M_4^2 (M_4^2\!-\!M_3^2)\!-\!2 M_3^2 (M_1^2\!-\!M_4^2) \right) 
       M_4^2 \ln\frac{M_4^2}{M_3^2}
\right.
\nonumber \\
\left. \frac{}{}
-(M_1^2-M_3^2) (M_4^2-M_3^2) (M_1^2-M_4^2) 
\left( M_1^2 (M_4^2-M_3^2)+2 M_3^2 (M_1^2-M_4^2) \right) \right]
\nonumber \\
+2 (M_4^2-M_3^2) (M_1^2-M_4^2) m_5^2 \ln\frac{m_5^2}{M_3^2}
\hspace{86mm}
\nonumber \\
\times\!
\left[ (M_1^2\!-\!M_3^2)^3 M_4^4 \ln\frac{M_4^2}{M_3^2}
\!-\!(M_4^2\!-\!M_3^2)^2 
\left( M_1^2 (M_1^2\!-\!M_3^2)\!+\!2 M_3^2 (M_1^2\!-\!M_4^2) \right) 
M_1^2 \ln\frac{M_1^2}{M_3^2}
\right.
\nonumber \\
\left. \frac{}{}
+(M_1^2-M_3^2) (M_4^2-M_3^2) (M_1^2-M_4^2) 
\left( M_4^2 (M_1^2-M_3^2)-2 M_3^2 (M_1^2-M_4^2) \right) \right]
\nonumber \\
-M_1^2 (M_4^2-M_3^2)^3
\ln^2\frac{M_1^2}{M_3^2} 
\left[ M_1^2 (M_1^2-M_3^2) (M_1^2-M_4^2) (k^2+m_2^2-m_5^2)
\right.
\hspace{26mm}
\nonumber \\
\left.
+2 M_4^2 (M_1^2-M_3^2)^2 k^2+2 M_3^2 (M_1^2-M_4^2)^2 m_2^2 \right]
\nonumber \\
-M_4^2 (M_1^2-M_3^2)^3
\ln^2\frac{M_4^2}{M_3^2} 
\left[ M_4^2 (M_4^2-M_3^2) (M_1^2-M_4^2) (k^2-m_2^2+m_5^2)
\right.
\hspace{26mm}
\nonumber \\
\left.
-2 M_1^2 (M_4^2-M_3^2)^2 k^2-2 M_3^2 (M_1^2-M_4^2)^2 m_5^2 \right]
\nonumber \\
- M_3^2 (M_1^2-M_4^2)^3
\ln\frac{M_1^2}{M_3^2} \ln\frac{M_4^2}{M_3^2} 
\left[ M_3^2 (M_1^2-M_3^2) (M_4^2-M_3^2) (k^2-m_2^2-m_5^2)
\right.
\hspace{16.5mm}
\nonumber \\
\left.
-2 M_1^2 (M_4^2-M_3^2)^2 m_2^2-2 M_4^2 (M_1^2-M_3^2)^2 m_5^2 \right]
\nonumber \\
+ (M_4^2\!-\!M_3^2) (M_1^2\!-\!M_4^2)
\ln\frac{M_1^2}{M_3^2} 
\left[  
M_3^2 (M_4^2-M_3^2) (k^2-m_2^2+m_5^2)
\right.
\hspace{40mm}
\nonumber \\
\times
\left( (M_1^2-M_4^2) 
\left( 2 M_1^2 (M_1^2-M_4^2)+M_1^4-M_3^4 \right)
       -2 (M_1^2+M_4^2) (M_1^2-M_3^2)^2 \right) 
\nonumber \\
-(M_4^2-M_3^2)^2 \left( 8 M_1^2 M_3^2 (M_1^2-M_4^2)
+3 M_1^2 (M_1^4-M_3^4) -M_4^2 (M_1^2-M_3^2)^2 \right) m_2^2
\hspace{10mm}
\nonumber \\
\left.
+(M_1^2\!-\! M_3^2) 
\left( 2 M_1^2 (M_4^2\!-\!M_3^2)^2 (M_4^2\!+ \! 2 M_3^2)
 \!-\!(M_1^2\!- \! M_3^2)^2 (M_1^2\!- \! M_4^2) 
  (M_4^2\!+\! M_3^2) \right) m_5^2 \right]
\nonumber \\
+ (M_1^2\!-\!M_3^2) (M_1^2\!-\!M_4^2)
\ln\frac{M_4^2}{M_3^2} 
\left[  
M_3^2 (M_1^2-M_3^2) (k^2+m_2^2-m_5^2)
\right.
\hspace{40mm}
\nonumber \\
\times
\left( (M_1^2-M_4^2) 
\left( 2 M_4^2 (M_1^2-M_4^2)-M_4^4+M_3^4 \right)
-2 (M_1^2+M_4^2) (M_4^2-M_3^2)^2 \right)
\nonumber \\
+(M_1^2-M_3^2)^2 
\left( 8 M_4^2 M_3^2 (M_1^2-M_4^2)-3 M_4^2 (M_4^4-M_3^4)
                        +M_1^2 (M_4^2-M_3^2)^2 \right) m_5^2
\hspace{10mm}
\nonumber \\
\left.
+(M_4^2\!-\!M_3^2) 
\left( 2 M_4^2 (M_1^2\!-\!M_3^2)^2 (M_1^2\!+\! 2 M_3^2)
           \!+\!(M_4^2\!-\!M_3^2)^2 (M_1^2\!-\! M_4^2) 
            (M_1^2\!+ \! M_3^2) \right) m_2^2 
\right]
\nonumber \\
-(M_1^2\!-\!M_3^2) (M_4^2\!-\!M_3^2) (M_1^2\!-\!M_4^2)^2
\left[ (M_1^2-M_4^2) (M_1^2+M_3^2) (M_4^2+M_3^2) 
(k^2\!-\!m_2^2\!-\!m_5^2)
\right.
\nonumber \\
-4 M_3^2 (M_1^2\!-\!M_4^2) \left( M_1^2 m_2^2 \!+\! M_4^2 m_5^2
+M_3^2 (k^2\!-\!2m_2^2\!-\!2m_5^2) \right)
\nonumber \\
\left. \left.
-2 (M_1^2-M_3^2) (M_4^2-M_3^2) (M_1^2 m_2^2 - M_4^2 m_5^2) 
\right] 
\frac{}{} \right\} .
\eea



\begin{thebibliography}{99}

\bbibitem{bardin}
  {\it ``Reports of the working group on precision calculations
  for the Z resonance''},
  D.Yu.~Bardin, W.~Hollik, G.~Passarino, eds. , CERN 95-03.

\bbibitem{Weigl} G.~Weiglein, R.~Scharf and M.~B\"{o}hm,
{\em Nucl.~Phys.} B416 (1994) 606; \\
D.~Kreimer, {\em Mod. Phys. Lett.} A9 (1994) 1105.

\bbibitem{sch} R.~Scharf,
Diploma Thesis, W\"{u}rzburg, 1991; Doctoral Thesis, W\"urzburg, 1994.

\bbibitem{berends1} F.A.~Berends, M.~B\"{o}hm, M.~Buza and R.~Scharf,
   {\em Z.~Phys.} C63 (1994) 227.

\bbibitem{GvdB} A.~Ghinculov and J.J. van der Bij, {\em Nucl. Phys.}
   B436 (1995) 30.

\bbibitem{Lunev} F.A.~Lunev, {\em Phys. Rev.} D50 (1994) 7735.

\bbibitem{Kreimer} D.~Kreimer, {\em Phys.~Lett.} B273 (1991) 277.

\bbibitem{CzKK} A.~Czarnecki, U.~Kilian and D.~Kreimer,
{\em Nucl. Phys.} B433 (1995) 259.

\bbibitem{berends2} F.A.~Berends and J.B.~Tausk,
{\em Nucl.~Phys.} B421 (1994) 456.

\bbibitem{bauberger} S.~Bauberger, F.A.~Berends, M.~B\"{o}hm 
and M.~Buza, {\em Nucl.~Phys.} B434 (1995) 383;\\
S.~Bauberger and M.~B\"{o}hm, {\em Nucl.~Phys.} B445 (1995) 25.

\bbibitem{japan} J.~Fujimoto, Y.~Shimizu, K.~Kato and Y.~Oyanagi, 
   KEK preprint 92-213.

\bbibitem{as-ex}
F.V.~Tkachov, Preprint INR P-358 (Moscow, 1984); \\
G.B.~Pivovarov and F.V.~Tkachov, Preprints INR P-0370, $\Pi$-459
   (Moscow, 1984); \\
K.G.~Chetyrkin, {\em Teor.Mat.Fiz.} 75 (1988) 26; 76 (1988) 207; \\
                      Preprint MPI-PAE/PTh 13/91 (Munich, 1991);\\
S.G.~Gorishny, {\em Nucl. Phys.} B319 (1989) 633;\\ 
V.A.~Smirnov, {\em Commun. Math. Phys.} 134 (1990) 109.

\bbibitem{Smirnov} V.A.~Smirnov,
{\em Renormalization and asymptotic expansions} 
(Birkh\"{a}user, Basel, 1991); \\
{\em Mod. Phys. Lett.} A10 (1995) 1485.

\bibitem{appl}
K.G.~Chetyrkin and A.~Kwiatkowski, {\em Phys.~Lett.}  B305 (1993) 285;
B319 (1993) 307; \\
K.G.~Chetyrkin, {\em Phys.~Lett.}  B307  (1993)  169; \\
S.A.~Larin, T. van Ritbergen and J.A.M.~Vermaseren,
{\em Phys.  Lett.}  B320  (1994) 159; B362 (1995) 134;
{\em Nucl.~Phys.} B438 (1995) 278;  \\
K.G.~Chetyrkin and O.V.~Tarasov, {\em Phys.~Lett.} B327 (1994) 114;\\
K.G.~Chetyrkin, J.H. K\"uhn and M.~Steinhauser,
{\em Phys. Rev. Lett.} 75 (1995) 3394.

\bbibitem{DT1} A.I.~Davydychev and J.B.~Tausk,
{\em Nucl.~Phys.} B397 (1993) 123.

\bbibitem{DST} A.I.~Davydychev, V.A.~Smirnov and J.B.~Tausk,
{\em Nucl.~Phys.} B410 (1993) 325.

\bbibitem{BDST} F.A.~Berends, A.I.~Davydychev, V.A.~Smirnov
 and J.B.~Tausk, {\em Nucl.~Phys.} B439 (1995) 536.

\bbibitem{FT} J.~Fleischer and O.V.~Tarasov,
{\em Z.~Phys.} C64 (1994) 413.

\bbibitem{special} D.J.~Broadhurst,
{\em Z.~Phys.} C47 (1990) 115; \\
D.J.~Broadhurst, J.~Fleischer and O.V.~Tarasov,
{\em Z.~Phys.} C60 (1993) 287; \\
R.~Scharf and J.B.~Tausk,
{\em Nucl.~Phys.} B412 (1994) 523.

\bbibitem{Tbilisi} D.T.~Gegelia, G.Sh.~Japaridze and 
K.Sh.~Turashvili, {\em Teor. Mat. Fiz.} 101 (1994) 225 
[{\em Theor. Math. Phys.} 101 (1994) 1313 ].

\bbibitem{dimreg} G.~'tHooft and M.~Veltman,
    {\em Nucl.~Phys.} B44 (1972) 189;\\
    C.G.~Bollini and J.J.~Giambiagi, {\em Nuovo Cim.} 12B (1972) 20.

\bbibitem{Reduce} A.C.~Hearn, {\em REDUCE user's manual},
      RAND publication CP78 (Santa Monica, 1987).

\bbibitem{'tHV'79} G.~'tHooft and M.~Veltman,
    {\em Nucl.~Phys.} B153 (1979) 365.

\bbibitem{BD} E.E.~Boos and A.I.~Davydychev, {\em Teor. Mat. Fiz.}
      89 (1991) 56 [ {\em Theor. Math. Phys.} 89 (1991) 1052 ]; \\
A.I.~Davydychev, {\em J. Math. Phys.} 32 (1991) 1052; 33 (1992) 358.

\bbibitem{ibp} F.V.~Tkachov, {\em Phys.~Lett.} B100 (1981) 65; \\
K.G.~Chetyrkin and F.V.~Tkachov, 
{\em Nucl.~Phys.} B192 (1981) 159; \\
F.V.~Tkachov, {\em Teor.~Mat.~Fiz.} 56 (1983) 350.

\bbibitem{JPA} A.I.~Davydychev, {\em J.~Phys.} A25 (1992) 5587.

\bbibitem{ChKT} K.G.~Chetyrkin, A.L.~Kataev and F.V.~Tkachov,
      {\em Nucl.~Phys.} B174 (1980) 345.

\bbibitem{DT3} A.I.~Davydychev and J.B.~Tausk, Mainz/Bergen preprint
     MZ-TH--95-26, Bergen--1995-14 (hep-ph/9511261);
     {\em Nucl. Phys.} B, to appear.

\end{thebibliography}
\end{document}